\input epsf
\documentstyle[preprint,aps,draft]{revtex}
\def\be{\begin{equation}}
\def\ee{\end{equation}}

\def\bea{\begin{eqnarray}   }
\def\eea{\end{eqnarray}   }

\def\r#1{(\ref{#1})}

\def\ie{{\it i.e.}}

\def\half{{\scriptstyle \frac{1}{2}}}
\def\quarter{{\scriptstyle \frac{1}{4}}}

\def\bra#1{\langle #1 |}
\def\ket#1{| #1 \rangle}
\def\bracket#1#2{ \langle #1 | #2 \rangle}

\def\bea{\begin{eqnarray}}
\def\eea{\end{eqnarray}}
\def\dee{\partial}

\def\deebydee#1#2{ \frac{\dee #1}{\dee #2}}

\tighten
\begin{document}

\preprint{SWAT/50/95, SUSX-TH-95/32}

\title{Quantum Dynamics Beyond the Gaussian Approximation}

\author{Gareth J. Cheetham}
\address{Department of Theoretical Physics,\\University College of Swansea,
\\Singleton Park,\\Swansea SA2 8PP\\{\tt G.Cheetham@swan.ac.uk}}
\author{E.J. Copeland}
\address{School of Mathematical and Physical Sciences,\\University of Sussex,\\
Brighton, BN1 9QH\\{\tt E.Copeland@susx.ac.uk}}
\date{March 23, 1995}
\maketitle


\begin{abstract}

The time dependent quantum variational principle is emerging as an
important means of studying quantum dynamics, particularly in early
universe scenarios. To date all investigations have worked within a
Gaussian framework. Here we present an improved method which is
demonstrated to be superior to the Gaussian approach and may be
naturally extended to the field--theoretic case.

\end{abstract}

\section{Introduction}


Early Universe scenarios tend to be based on the evolution of scalar
fields, either through their r\^ole as inflaton fields or as
topological defect forming fields.  A detailed understanding of the
quantum evolution of these fields is important in order to fully
describe their behaviour during a phase transition.  This has recently
been the focus of a great deal of attention. Its study relies on
analysing the quantum dynamics in real time.  Guth and Pi
\cite{GUTHPI} in one of the land mark papers in the field of inflation
investigate the quantum mechanics of the scalar field in the new
inflationary universe, looking in detail at the ``slow rollover''
transition.  The idea of a ``slow rollover'' arises because the
transition involves a scalar field $\phi$ which evolves slowly down
its potential starting from some initial position where it is
described by a well defined wave function. The phase transition can be
thought of one where at very high temperatures the potential has a
minimum at $\phi =0$, which becomes unstable as the temperature
decreases, with the stable minima moving to a new larger value of
$\phi = \pm \sigma$ say.  As the universe cools the field remains
close to $\phi=0$, slowly evolving towards its true vacuum value. The
beginning of the phase transition is quantum mechanical in nature, yet
the late time evolution of the scalar field being described by
classical equations of motion. This assertion needs to be justified,
as first addressed in \cite{GUTHPI}.  Analysing an exactly soluble
linearised model, both in one-dimensional quantum mechanics and in
quantum field theory of a single scalar field, it was discovered that
the large-time behaviour of the field in an unstable upside down
harmonic oscillator potential is `accurately described by ``classical
physics.'''  In the one-dimensional quantum-mechanical model the
evolution of the wave function describing the particle in the
potential is determined by the exactly solvable Schr\"odinger
equation. The solution for the wave function is not surprisingly that
of a Gaussian. The harmonic oscillator potential maintains the form of
the initial Gaussian wave function.

In \cite{COOPER} the time--dependent variational method developed by
Jackiw and Kerman \cite{JACKIW} is used in the investigation of the
behaviour of a particle moving in a one dimensional quantum mechanical
with more realistic potentials (i.e. a double well potential), which
have analogues both in the inflationary universe scenario and in
models for the formation of topological defects.  The important point
that needs to be raised, and the motivation for this paper, is that in
that work, and most subsequent work, the analysis is performed using a
Gaussian trial wavefunction. The resultant equations of motion
obtained through the variation of the effective action are the
time-dependent Hartree-Fock (HF) equations.  In \cite{COOPER} the
authors argue that by comparison with the exact (numerical) solution
the variational HF approximation accurately describes the process, and
that the late time behaviour of the evolution is approximately
classical if described in terms of a suitably chosen small
dimensionless coupling constant.  We argue in this paper, that it is
straightforward to go beyond this Gaussian ansatz by expanding the
wavefunction in a complete set of Hermite polynomials. In particular
we find that just keeping the first and second order terms in the
expansion leads to a dramatic improvement in the accuracy of the
variational approach, and argue that such a technique can easily be
adapted to the field theory \cite{JACKIW89} case where presently the
Hartree- Fock approximation is generally adopted in variational
calculations applied to the early Universe \cite{DEVEGA}.

There are many reasons why it would be advantageous to go beyond the
Gaussian approximation. If we have a potential which has degenerate
minima, then it is impossible for a Gaussian wavepacket to accurately
describe the evolution of a scalar field during a phase
transition. This is particularly relevant for calculations involving
the formation of topological defects. For example imagine we wish to
understand the circumstances under which defects can be said to have
formed and their distribution at formation\cite{KIBBLE}.  Recently in
the context of $^{4}$He vortices, there has been considerable
attention paid to understanding the evolution of the scalar field
responsible for their formation just after the quench transition
\cite{RIVERS}.  One of the limitations of this interesting calculation
is that it can not accurately probe the non--linear regions of the
potential, hence is only strictly valid just after the quench.  In
order to fully describe the formation process it is important to be
able to probe the true vacuum of the potential.

In the context of our simple one dimensional quantum mechanical system
we hope to demonstrate that by extending the ansatz of the wavefunction
to include the Hermite polynomials, then at little extra cost in
complexity we can probe the non--linear region of the potential in far
greater detail than has previously been possible; an advantage that will
be carried over to the field--theoretic treatment.

Another area where the non--linearities of the theory need to be
probed, is in the reheating calculations associated with inflation. As
the inflaton field evolves down its potential, eventually it moves out
of the ``slow roll'' regime as it descends into the true minima of the
potential. The traditional picture is that in this region as the field
oscillates about this minima then it decays through coherent
oscillations and reheats the universe, restoring the radiation
dominated universe. Recently though, this picture has been questioned
\cite{LINDE94,HOLMAN}. In order to fully probe this region of the
potential, it is important that the ansatz adopted for the scalar
field is valid in this region.  It is our belief that the usual
Hartree-Fock approximations are not sufficient here and need to be
improved.  The method we outline in the rest of this paper is one
possible way of improving the situation.


Use of the time--dependent variational principle of quantum mechanics
is becoming more widespread. The method, first given by Dirac,
involves the construction of an `effective action'\footnote{The
connection of which to the usual effective action, or Gibbs free
energy is given in \cite{JACKIW}. }
\be
\Gamma=\int dt \bra{\psi} i\dee_t-\hat{H}\ket{\psi},
\ee
where $\dee_t \equiv \partial / \partial t$ and $\hat{H}$ is the
Hamiltonian operator. $\Gamma$ is then made stationary $\delta
\Gamma=0$ against variations of the state $\bra{\psi}$ subject to the
constraint $\bracket{\psi}{\psi}=1$. Approximate\footnote{It is not
actually clear in what way the dynamics so obtained are approximate,
an issue we will address in a later paper.} dynamics are obtained by
positing a variational ansatz for the wavefunction which is a
function of a small number of variables.

Central to the approach is the assumption that the ansatz for
the variational wavefunction is `close' to the exact one,
\ie\ there are sufficient degrees of freedom for the
wavefunction to accurately track the evolution of the system.

To date investigations have been restricted to the use of Gaussian
ansatzes since these are calculationally easy to handle. However
we shall argue that the Gaussian approach is of limited applicability
and that results gained from it have a limited range of reliability.

\section{The Variational Procedure in Generality}
\label{GENERAL}

In order to elucidate our later calculations it is useful to consider
this construction in generality. Consider the variational effective action
\be
\Gamma=\int dt \bra{\psi} i\dee_t-\hat{H}\ket{\psi}= \int dt
\frac{i}{2}\left( \bra{\psi}\dee_t\ket{\psi}-
(\dee_t\bra{\psi})\ket{\psi}\right) -\bra{\psi}\hat{H}\ket{\psi}. \ee
Let us suppose the variational state to be a function of $n$ real
parameters $v_i$
\be
\Gamma=\int dt \frac{i}{2}\left( \bra{\psi}\deebydee{\ket{\psi}}{v_i} -
\deebydee{\bra{\psi}}{v_i}\ket{\psi}\right)\dot{v}_i-\bra{\psi}\hat{H}\ket{\psi}.
\ee
When the action is made stationary with respect to variation of these
parameters we obtain the induced equations of motion
\be
i\left[ \deebydee{\bra{\psi}}{v_j}\deebydee{\ket{\psi}}{v_i} -
(i\leftrightarrow j)\right]\dot{v}_i
-\deebydee{}{v_j}\bra{\psi}\hat{H}\ket{\psi}=0.
\label{EOMS}
\ee
Schematically, this expression is of the form
\be
A_{ij}\dot{v}_j-b_i=0
\ee
implying that in order that we are able to extract the equations of motion
for the parameter $v_j$, the
matrix $A$ must be non--singular throughout the evolution of the system.

Obviously thus far we have simply expressed the process one performs
in arriving at the approximate equations of motion. However, given in
this form, the equations of motion are simple to extract and the
origin of possible singularities is highlighted.

\section{Improved Wavefunction}

Before we give the expression for the improved wavefunction let us
make the definition
\be
u_n(x) :=\left( \frac{\alpha}{\pi^\half 2^n n!}\right)^\half H_n(\alpha x)
e^{-\alpha^2 x^2/2} \quad \alpha=(2G)^{-\half}.
\ee
where $G(t)$ is real and $H_n$ is the $n$th Hermite polynomial.
We see that the $u_n$ are a one parameter set of orthonormalized functions
\be
\int u_n(x) u_m(x) dx =\delta_{nm}
\ee
independent of the value of $G$. We may thus use the $u_n$ as a basis
for our variational wavefunction:
\be
\psi(x,t)=N e^{i\Pi x^2}\sum_{n=0}^\infty a_n u_n(x,t).
\ee
Here the $a_n$ are time dependent complex numbers, the other real variational
parameters being $\Pi$ and $G$ which implicitly appear in the definition
of the $u_n$.
Normalization is achieved by the inclusion of a time dependent $N$.

As it stands this represents no simplification. Our plan of
action is therefore to truncate the expansion at some finite order and
work consistently to that order. The zeroth order approximation
is simply the Gaussian, Hartree--Fock approximation.

As a test bed for our method we will consider the case of a particle
moving in the potential considered by Cooper et al. \cite{COOPER}, as
this illustrates well the shortfalls of the Gaussian approach. In a
later paper the method will be applied to a wider range of systems.

The potential we consider is of a double well,
\be
V(x)=\frac{\lambda}{24}(x^2-a^2)^2
\ee
with Gaussian initial conditions $G_0=\sqrt{3/2\lambda a^2}$, and
$a$ is the symmetry breaking value for $x$.

To demonstration of the power of the method we will only include the
first non--trivial term in the expansion. Since the potential and
initial conditions are symmetric, the first non--trivial term involves
$u_2$
\be
\psi=N e^{i\Pi x^2}(u_0(x,t)+a_2 u_2(x,t)).
\ee
Even to this order we shall see that the improvement in the results
over those of the Gaussian approach is impressive.
We shall compare the results obtained using the improved equations of
motion with the exact results obtained via numerical simulation.

\section{Equations of motion}

Working with a polar representation of $a_2(t)=R
e^{i\theta}$, the equations of motion one obtains from \r{EOMS} are
\bea
\dot{G} & = & 4\Pi G -\frac{\sqrt{2}G^3 s \lambda}{6 R} \\
\dot{\Pi} & = & \frac{1}{8 G^2} -2\Pi^2+\frac{\lambda
a^2}{12}-\frac{7G\lambda}{12} -\frac{\lambda \sqrt{2}cG}{24 R} \\
\dot{R} & = & s \lambda G^2\frac{(c + R^2 c
+2R\sqrt{2}+2R^3\sqrt{2})}{6R} \\
\dot{\theta} & = & -\frac{\lambda
G^2(4R^3\sqrt{2}c+2c^2R^3-2c^2+1-6R\sqrt{2}c-11R^2)}{12 R^2}
-\frac{1}{G}
\eea
where $s = \sin \theta$ and $c = \cos \theta$.
The increase in complexity of the result over that of the Gaussian
approach is more than compensated for by the increase in the accuracy of
the results.

We notice that the improved equations of motion have within them terms
familiar from the Gaussian approach. However, the HF equations of
motion are not obtainable as a simple limit $a_2\rightarrow 0$ as the
improved equations are singular in this limit. The origin of this
singularity is as was outlined in section \r{GENERAL}.

To assess the use of method we shall focus on the evolution of the
quantity
\be
\langle \hat{x}^2 \rangle = \frac{G(1+5R^2+2\sqrt{2}Rc)}{1+R^2}.
\ee

\section{Initial Conditions}

It is clear that the equations of motion are singular for the Gaussian
initial conditions considered; since our aim is a comparison of the
results of the improved method with those of Cooper et al, we shall
adopt the same initial conditions. This apparent problem is easily
circumvented.

The rationale is to start the evolution of the system some short time
after $t=0$. We solve the exact Schr\"odinger equation to first order
and may then extract the values of $R$ and $\theta$, using these as
the initial conditions in the variational equations of motion. In
actuality the subsequent evolution of the system is rather insensitive
to the initial conditions.

Following \cite{COOPER}, we take as the initial
conditions
\be
\psi_0(x)=(2\pi G_0)^{-\quarter}e^{-x^2/4G_0}
\ee
where $G_0=\sqrt{\frac{3}{2\lambda a^2}}.$

After a short time $\delta t$ this evolves into
\be
\psi(x,\delta t)=\psi_0(x)-i\delta t \hat{H} \psi_0(x)
\ee
which is the state we match onto. This procedure leads us to take
\be
R_0=\frac{\delta t\sqrt{2}(6+\lambda a^4 G_0)}{24} \quad
\theta_0=\frac{\pi}{2}.
\ee

\section{Results and Conclusions}

\begin{figure}
\epsfbox{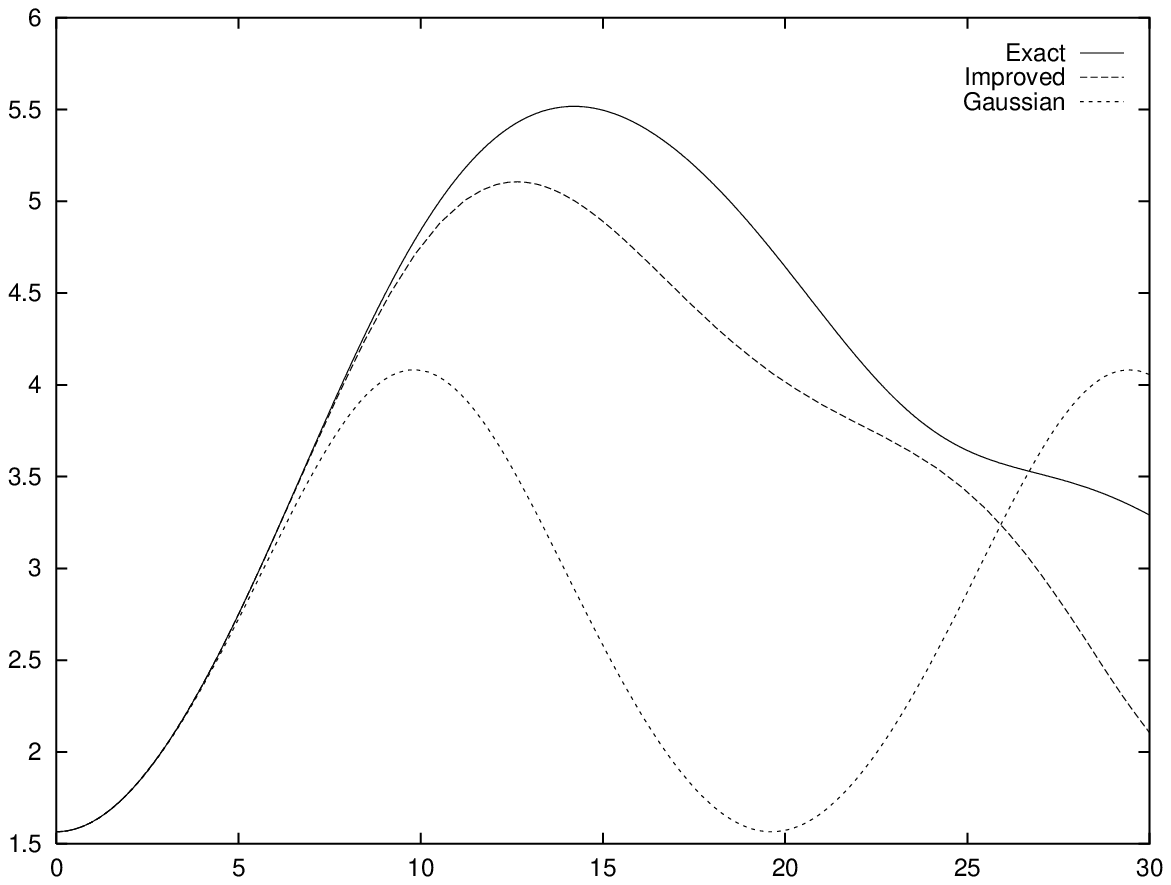}
\label{FIG1}
\caption{Comparison of the methods for $a=5$.}
\end{figure}

\begin{figure}
\epsfbox{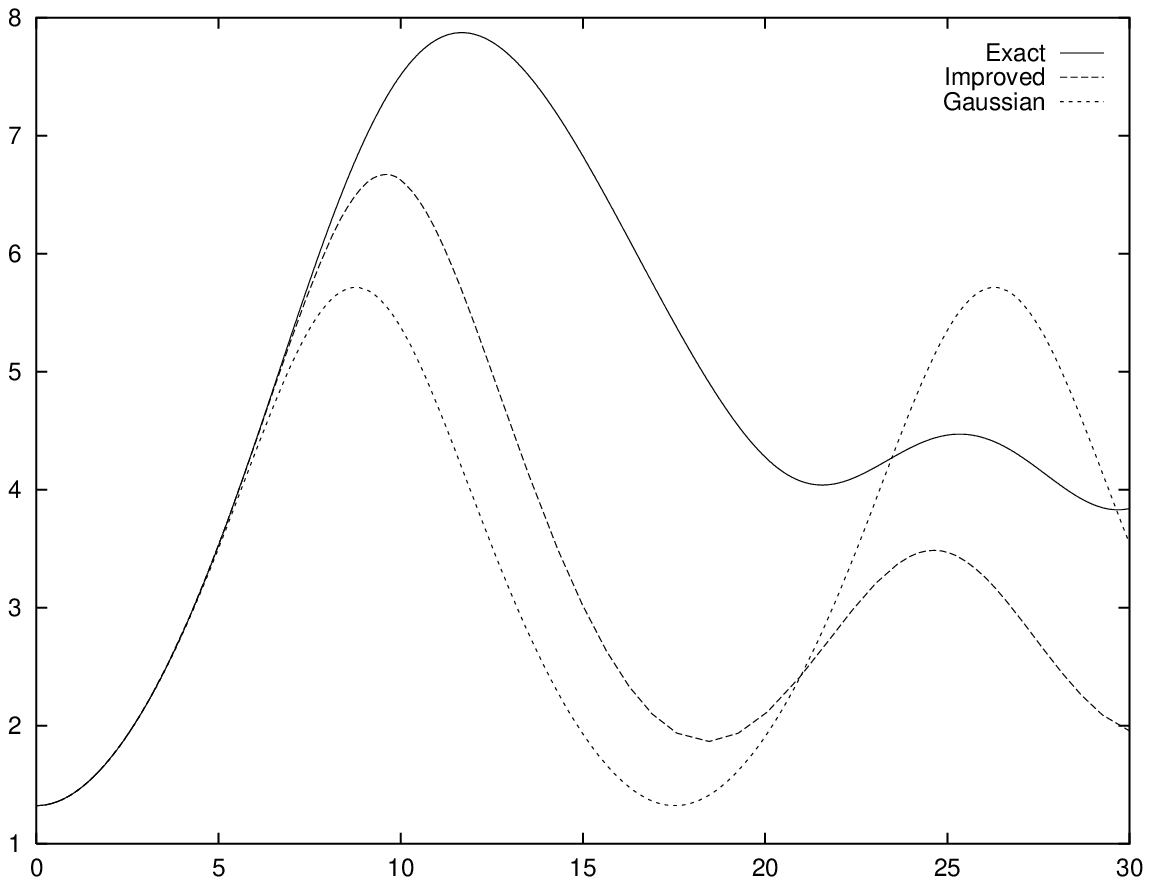}
\label{FIG2}
\caption{Comparison of the methods for $a=7$.}
\end{figure}

The results of the above calculation are presented  in figures
\ref{FIG1} and \ref{FIG2}. Plotted are the exact evolution of
$\langle{x^2}\rangle$, found by numerical simulation, against the
improved and Gaussian results for the two values $a=5,7$. It is clear
that the improved method furnishes us with a result considerably
closer to the exact evolution than does the Gaussian. Also we see that
the improved method samples regions of the potential much closer to
the minima than does the HF, the so--called spinodal regions,
suggesting that more information about the potential is being taken
into account.  The Gaussian wavefunction leads to a turning point of
$\langle x^2 \rangle$ at $2/3 a^2$
\cite{COOPER}, which provides an indication of where the ansatz breaks down.
With the improved ansatz we find the turning point occurs typically at
$\langle x^2\rangle\sim a^2$, demonstrating the significant
increase in accuracy. Moreover, this approximate solution clearly
probes the non-linear region of the potential.

Behind this success is the crucial observation that the improved
wavefunctions are capable of becoming bimodal in nature, something not
open to the Gaussian ansatz. This means that we have a method of
investigating the field evolution during a defect forming transition
and a first order transition. These cases are currently being analysed.

Perhaps the most promising aspect of this work is the possible
extension to the field theoretic case. We may identify the Hermite
polynomials as the eigen--solutions of the harmonic oscillator. In
field theory the analogous system is the massive free--field. The
eigensolutions for this system are easily calculable and may then be
used as a basis, as above. This work is on--going.

\end{document}